\documentclass[twocolumn,amsmath,amssymb,prb,superscriptaddress, achemso, altaffilsymbol, footinbib]{revtex4-1}
\pdfoutput=1
\makeatletter
\renewcommand*{\@fnsymbol}[1]{\ensuremath{\ifcase#1\or \dagger\or *\or \ddagger\or
   \mathsection\or \mathparagraph\or \|\or **\or \dagger\dagger
   \or \ddagger\ddagger \else\@ctrerr\fi}}
\makeatother

\usepackage{graphicx,amsmath, amssymb, pifont}
\usepackage[usenames]{color}
\usepackage{pifont}
\usepackage{gensymb}
\usepackage{wasysym}

\begin{document}

\preprint{}

\title{Bottom-up assembly of metallic germanium}
\author{Giordano Scappucci}
 \email{giordano.scappucci@unsw.edu.au}
\author{Wolfgang M. Klesse}
\author{LaReine A. Yeoh}
\affiliation{School of Physics, University of New South Wales, Sydney, 2052, Australia}
\author{Damien J. Carter}
\affiliation{Department of Chemistry, Curtin University, Perth WA 6845, Australia}
\affiliation{Nanochemistry Research Institute, Curtin University, Perth WA 6845,  Australia}
\author{Oliver Warschkow}
\affiliation{Centre for Quantum Computation and Communication Technology, School of Physics, The University of Sydney, Sydney NSW 2006, Australia}
\author{Nigel A. Marks}
\affiliation{Department of Physics and Astronomy, Curtin University, Perth WA 6845, Australia}
\affiliation{Nanochemistry Research Institute, Curtin University, Perth WA 6845, Australia}
\author{David L. Jaeger}
\affiliation{Department of Material Science and Engineering, University of North Texas, Denton, Texas 76209, United States}
\author{Giovanni Capellini}
\affiliation{IHP, Im Technologiepark 25, 15236 Frankfurt (Oder), Germany}
\affiliation{Dipartimento di Scienze, Universit\`a Roma Tre, Viale Marconi 446, 00146 Rome, Italy}
\author{Michelle Y. Simmons}
\affiliation{School of Physics, University of New South Wales, Sydney, 2052, Australia}
\affiliation{Centre of Excellence for Quantum Computation and Communication Technology, School of Physics, University of New South Wales, Sydney, New South Wales 2052, Australia}
\author{Alexander R. Hamilton}
\affiliation{School of Physics, University of New South Wales, Sydney, 2052, Australia}

\maketitle

\normalsize

\textbf{Extending chip performance beyond current limits of miniaturisation requires new materials and functionalities that integrate well with the silicon platform. Germanium fits these requirements and has been proposed as a high-mobility channel material,\cite{1} a light emitting medium in silicon-integrated lasers,\cite{2,3} and a plasmonic conductor for bio-sensing.\cite{4,5} Common to these diverse applications is the need for homogeneous, high electron densities in three-dimensions (3D). Here we use a bottom-up approach to demonstrate the 3D assembly of atomically sharp doping profiles in germanium by a repeated stacking of two-dimensional (2D) high-density phosphorus layers. This produces high-density (10$^{19}$ to 10$^{20}$~cm$^{-3}$) low-resistivity (10$^{-4}~\Omega\cdot$cm) metallic germanium of precisely defined thickness, beyond the capabilities of diffusion-based doping technologies.\cite{6} We demonstrate that free electrons from distinct 2D dopant layers coalesce into a homogeneous 3D conductor using anisotropic quantum interference measurements, atom probe tomography, and density functional theory.}
	
Doping germanium with homogeneous, free-electron concentrations well above the metal-insulator transition ($n_{3D}$ = 10$^{17}$~cm$^{-3}$) enables low-resistivity source/drain extensions in high mobility transistors,\cite{7} Ge-on-Si integrated lasers with maximal optical gain,\cite{8} and plasma wavelengths (2--5 um) suitable for biological sensing.\cite{5} Mainstream top-down implantation is inadequate for this purpose because vacancy-enhanced diffusion and the formation of neutral complexes create electrical deactivation and doping profiles that are broad and inhomogeneous.\cite{9,10} Self-limiting surface reactions provide a promising alternative to control doping processes from the bottom-up.\cite{11} This approach has produced monolayer-doped semiconductors with high-density, strongly confined, two-dimensional electron gases (2DEGs).\cite{12} Combinations of bottom-up and top-down approaches have been proposed\cite{13,14} to extend monolayer doping from 2D to 3D. While dopants are deposited in a single 2D layer, their distribution in 3D was obtained by thermal diffusion, with associated loss of atomic precision and profile homogeneity.

\begin{figure*}
\center
\includegraphics[width=170mm]{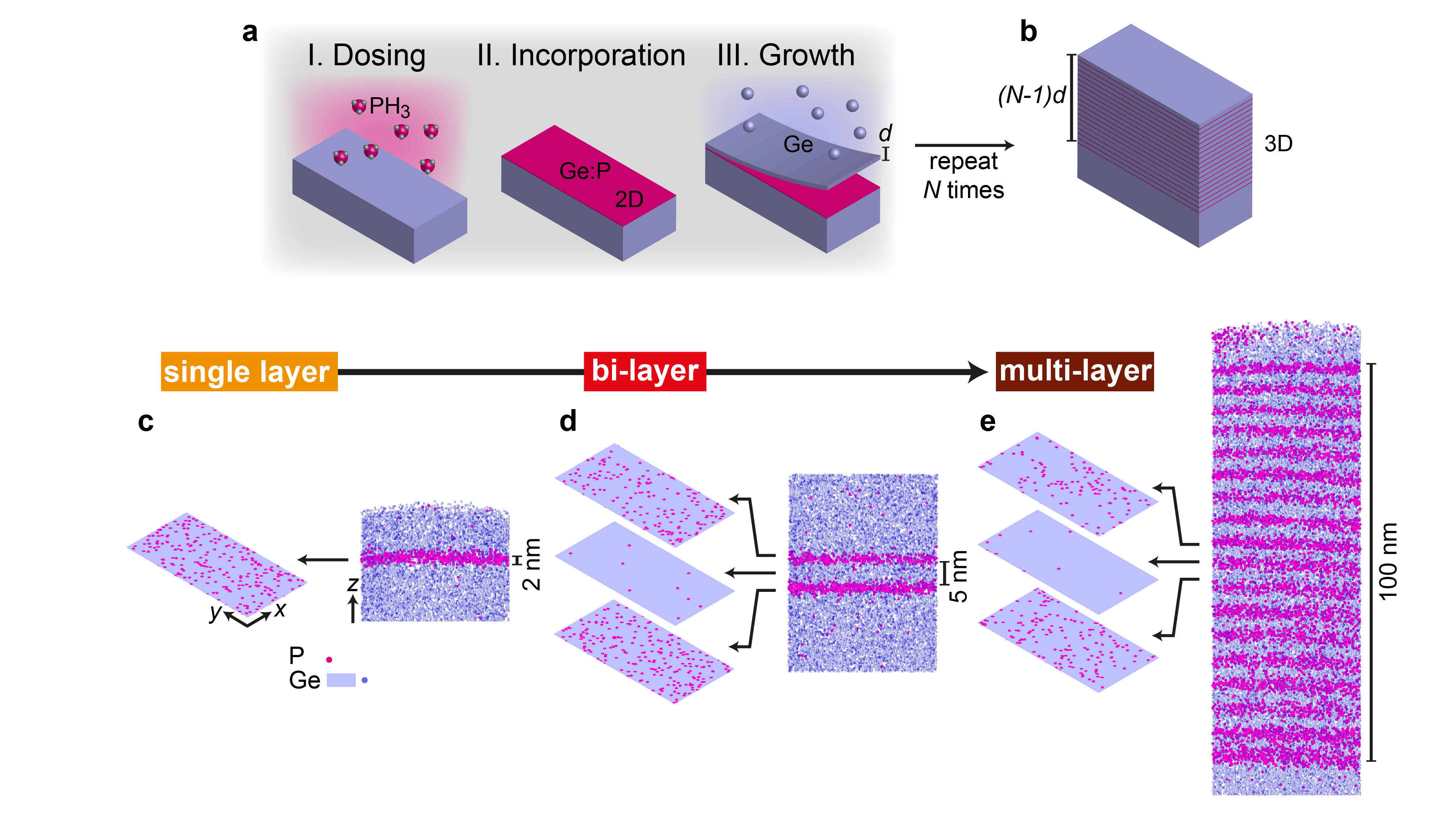}
\caption{\small  \label{fig1} \textbf{Three-dimensional assembly of atomically sharp doping profiles using a bottom-up approach}.(a) Phosphorus doped layers in germanium are fabricated in ultra-high vacuum by adsorption of phosphine molecules (PH$_{3}$) onto a clean Ge(001) surface, thermal incorporation of P atoms, and encapsulation under an epitaxial layer of germanium of thickness $d$. (b) Repetition of the sequence in (a) produces a highly doped Ge film of total thickness $(N$$-$$1)d$. Pulsed laser atom probe tomography results from a (c) single layer, (d) bi-layer, and (e) multi-layer (18 layers) samples showing the cross-section distribution of dopant atoms.}
\end{figure*}

	Here we demonstrate an exclusively bottom-up approach to produce an effectively 3D doped, high-density, low-resistivity metallic germanium with a precisely defined thickness and doping profile. This is achieved by the repeated deposition of $N$ nearly-identical phosphorus doped layers as illustrated in Fig.~1a. This approach preserves the vertical atomic-precision associated with monolayer doping and creates a homogeneous 3D system, provided the interlayer spacing is sufficiently small that the electrons can readily move between the layers. Each layer is prepared in a three-step process (Fig.~1a): (1) self-saturating chemisorption of phosphine molecules (PH$_{3}$) onto a clean Ge surface;\cite{15} (2) substitutional incorporation of P dopants into the Ge lattice by thermal annealing to provide a two-dimensional electron density $n_{2D}$;\cite{12} (3) encapsulation with Ge by molecular beam epitaxy to separate layers at a distance $d$. By stacking a large number of layers a precisely-doped slab of thickness $h = (N-1)d$ is produced. Using an inter-layer separation $d$ comparable to the Bohr radius of phosphorus in germanium ($\approx$8~nm), vertical electron delocalisation creates a homogeneous 3D system with a density $n_{3D} = Nn_{2D}/h$. With current monolayer doping techniques\cite{16} achieving $n_{2D}$ in the range of 10$^{13}$ to 10$^{14}$~cm$^{-2}$ and with sub-10~nm layer separations, we can expect high 3D densities $n_{3D}$ in the range of 10$^{19}$ to 10$^{20}$~cm$^{-3}$. The transition from single 2D layers to an effectively 3D doped material is explored in this paper using a single-layer (Fig.~1d), a bi-layer (Fig.~1e), and a multi-layer sample (Fig.~1f) grown in ultra-high vacuum using the sequence described in Fig.~1a (Methods section). We use an interlayer separation of $d\approx5.7$~nm, which is less than the Bohr radius.
	
	Three-dimensional atom probe tomography of these samples (Fig.~1c--e) shows dopant layers that are narrow and well separated. The dopant distribution of the single-layer sample has a full width at half maximum of $1.41\pm0.05$~nm. The P atoms are distributed randomly within the doping plane with a very high planar density of $1.44 \times 10^{14}$~cm$^{-2}$. A similar average width ($1.4\pm0.1$~nm) and density ($1.2\pm0.3\times10^{14}$~cm$^{-2}$) is found for layers of the bi-layer sample. In the multi-layer sample the average width and density per layer are $2.0\pm0.4$~nm and ($1.2\pm0.3)\times10^{14}$~cm$^{-2}$, respectively. There is a gradual increase in width by 0.06~nm from one layer to the next due to the accumulated thermal budget of the repeated deposition process (see Supplementary Section 1 for details). Crucially, we find that the inter-layer separation is preserved from layer to layer, averaging 5.65~nm with a variance of less than 0.05~nm. These metrics confirm that the vertical atomic-precision associated with monolayer doping is largely preserved when multiple layers are assembled into a larger stack.
	
	By comparing atom probe tomography and Hall measurements (see Methods) we determine that approximately 26 to 44\% of the dopants are electrically active. Despite incomplete activation, presumably due to the formation of P-P dimers,\cite{16} the measured electronic planar densities $Nn_{2D}$ are high at 6.3, 12, and $56\times10^{13}$~cm$^{-2}$ in the single layer, bi-layer and multi-layer (18-layer) sample, respectively. This corresponds to ultra-high 3D electronic densities of 4.5, 1.9, and $0.6\times10^{20}$~cm$^{-3}$. We measure exceptionally low resistivities of 2.0, 4.5, and $6.7\times10^{-4}$ $\Omega\cdot$cm that are consistent with the resistivity vs.\ density dependence expected in heavily-doped bulk Ge.\cite{17,18} This demonstrates that our bottom-up, “interface-free” doping technology achieves bulk-like resistivities in extremely thin ($\approx$1.4~nm) doping profiles.
		
\begin{figure*}
\center
\includegraphics[width=170mm]{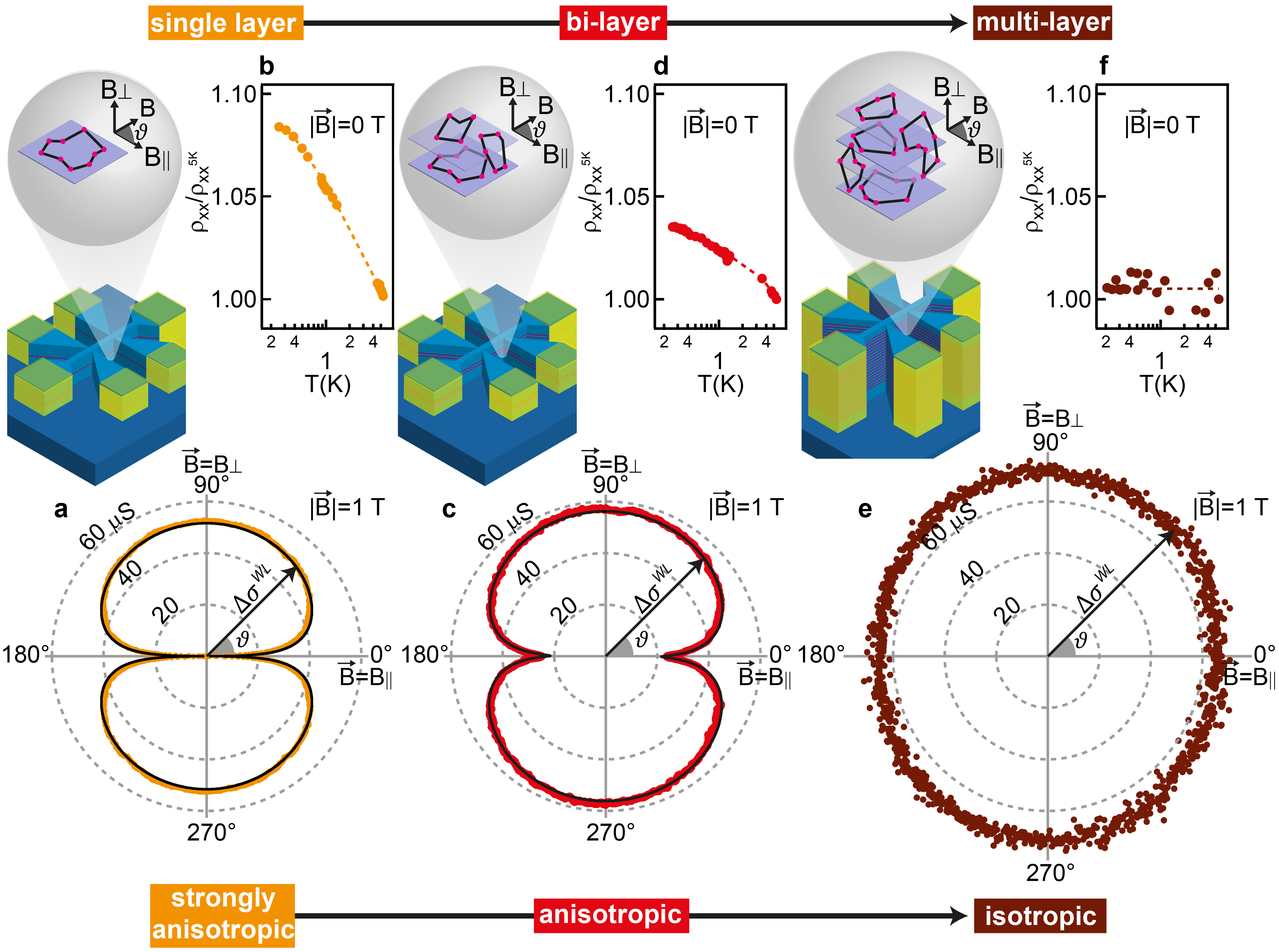}
\caption{\small  \label{fig2} \textbf{Evolution of the electronic system from 2D to 3D probed by electronic transport in different magnetic field orientations.}(a), (c), (e) Polar plots of the size of the weak localization positive corrections to the magnetoconductivity $\Delta\sigma^{WL}(\vartheta)=\sigma_{xx}(|\vec{B}|=1~\mbox{ T}, \vartheta)-\sigma_{xx}(|\vec{B}|=0~\mbox{ T})$ as a function of angle $\vartheta$ at a temperature of 200~mK for the single layer, bi-layer, and multi-layer samples, respectively; the angle $\vartheta$ is defined with respect to the dopant plane, with the components of magnetic field perpendicular and parallel to the plane of the dopant layers given by $B_{\perp}=|\vec{B}|\sin(\vartheta)$ and $B_{\lVert}=|\vec{B}|\cos(\vartheta)$. Dashed grey lines are contours of constant $\Delta\sigma^{WL}(\vartheta)$. Black lines are theoretical fits to the data;  (b), (d), (f) Temperature dependence of the zero magnetic field resistivity $\rho_{xx}$ (normalized to its value $\rho_{xx}^{5K}$ at $T=5$~K) measured for the same samples (lines are a guide for the eye), showing how the characteristic log$T$ behaviour for 2D systems weakens as the number of doping layers is increased. Illustrations of the Hall bars and weak localisation backscattered particle trajectories are also shown for the three samples.}
\end{figure*}

	Quantum interference measurements at cryogenic temperatures and in a vector magnetic field ($\vec{B}$) allow us to probe electron motion both within and between dopant layers, and observe the evolution from 2D towards a homogeneous 3D electronic system as the number of layers is increased. Figure 2 shows polar plots $\Delta\sigma^{WL}(\vartheta)$ of the weak localization (WL) positive corrections to the magnetoconductivity measured at constant field and variable angle $\vartheta$. The single layer sample (Fig.~2a) exhibits a strong anisotropy with $\Delta\sigma^{WL}(\vartheta)$  at a maximum when $\vec{B}$ is perpendicular to the dopant plane. As the magnetic field is rotated, $\Delta\sigma^{WL}(\vartheta)$  collapses and is nearly negligible when $\vec{B}$ is parallel to the 2D layer. This finding is intuitive: the 2DEG is strongly confined in the vertical direction and hence the backscattered particle trajectories that give rise to WL are located within the 2D dopant plane. In the bi-layer sample (Fig.~2c), the anisotropy is still present but diminished. This indicates that in addition to motion within the layers, electrons are able to jump between them and form coherence interference loops that are affected by the parallel magnetic field. The two layers are strongly coupled despite the layer separation and effectively behave as a coherent 2D system of finite thickness. Quantitative analysis of the weak localisation yields the time scales associated with elastic scattering ($\tau_{e}$), interlayer tunneling ($\tau_{t}$), and dephasing ($\tau_{\varphi}$). The analysis shows $\tau_{t}\leq\tau_{e}\ll\tau_{\varphi}$   (see Supplementary Section 2 for details), confirming coherent tunnelling of electrons between layers over a timescale comparable to scattering off dopants within each layer. In the multi-layer sample, the polar plot (Fig.~2e) is nearly circular. The electron self-intersecting scattering paths are equally probable in all the three directions, supporting the presence of a homogeneous 3D system.

	Temperature-dependent resistivity measurements at zero magnetic field provide additional confirmation of this 2D to 3D cross-over. Figures 2b, d, and f show how the log$T$ dependence characteristic of a 2D system\cite{19} is increasingly suppressed as layers are vertically stacked. Overall, this suppression of the temperature dependence and the progressive loss of anisotropy of the electron quantum interference reflect the loss of vertical confinement as electron motion approaches that of a 3D metallic conductor. Despite the atom probe imaging demonstrating that the sample comprises well-defined 2D layers, the strong inter-layer coupling means that the electrons see the sample as a bulk 3D material.

\begin{figure*}
\center
\includegraphics[width=170mm]{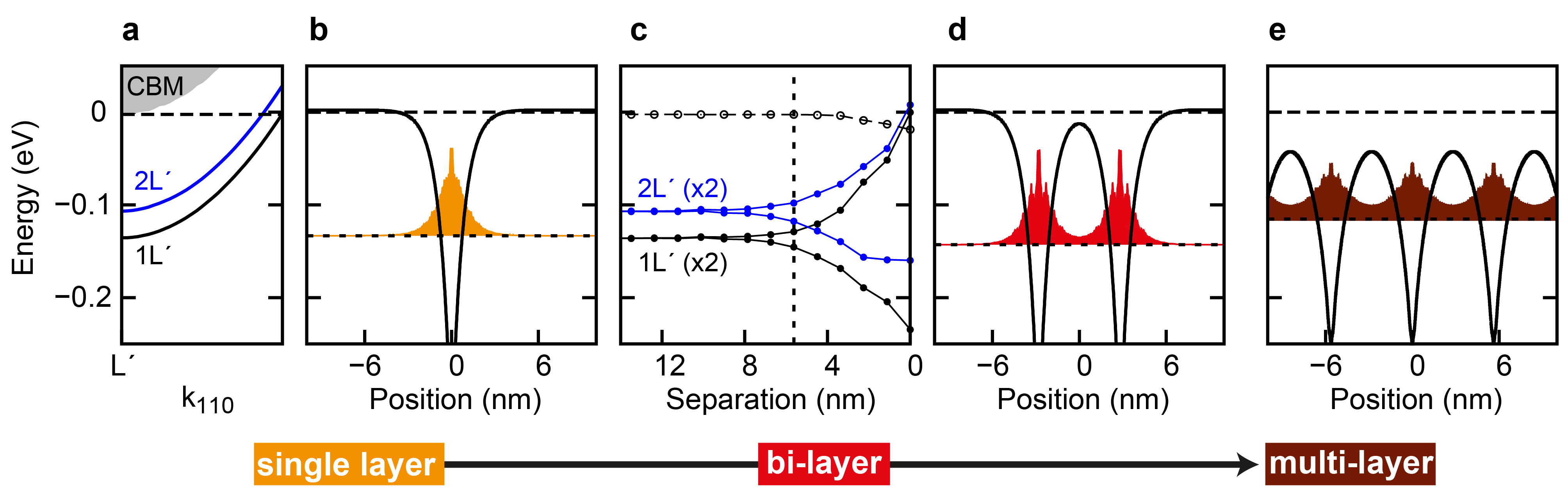}
\caption{\small  \label{fig3} \textbf{Density functional calculations of stacked dopant layers in germanium.} (a) Band structure of a single phosphorus dopant layer ($6.3\times10^{13}$~cm$^{-2}$) plotted from the 2DEG band minimum towards the zone centre. The bulk conduction band minimum (CBM) is indicated using grey shading. (b) Dopant potential and donor electron density for the same single layer structure. (c) Correlation of band energies for a pair of phosphorus dopant layers (both $6.3\times10^{13}$~cm$^{-2}$) as a function of separation. The Fermi energy is indicated by empty circles. For reference, the separation in the experimental bi-layer sample (5.7~nm) is indicated by a vertical dashed line.  (d) The dopant potential and donor electron density for the same pair of dopant layers at the experimental separation. (e) Dopant potential and donor electron density for an infinite stack of dopant layers (each $3.1\times10^{13}$~cm$^{-2}$) with a spacing of 5.7~nm.}
\end{figure*}

	The thickness $h$ of the conducting 2D systems in the single and bi-layer samples can be extracted by fitting $\Delta\sigma^{WL}(\vartheta)$ in Fig.~2a,~c to a generalized angle-dependent Hikami-Larkin-Nagaoka expression\cite{20,21} (see Supplementary Section 3 for more details) that includes an additional dephasing rate due to the parallel magnetic field.\cite{22,23} We obtain thicknesses of $h=1.49\pm0.03$ and $6.17\pm0.05$~nm for the single and bi-layer sample, respectively, in agreement with thicknesses of $1.41\pm0.05$ and $7.3\pm0.2$~nm obtained by atom probe tomography. This confirms the interpretation of the multi-layered doped region as a 3D space where electrons coming from 2D layered dopants can freely move, thus realizing a 3D metallic conductor.
	
	The critical role of the interlayer spacing $d$ on the electronic transition from 2D to 3D behaviour is further explored using density functional theory in which the activated dopant densities are used as input parameters.  A single phosphorus layer in germanium\cite{24} is characterized by a pair of valley-split bands (Fig.~3a, labelled 1L$^\prime$ and 2L$^\prime$), and a 2DEG density that is spatially confined to a width of $\approx7.3$~nm by the self-consistent doping potential (Fig.~3b). Figure 3c describes the evolution of the band minima of two dopant layers as the separation between the layers is reduced. At large separation, the two layers are effectively independent and the single layer band energies are preserved. At closer separation the band energies split due to coupling between the two layers. This commences at $d\approx10$~nm and is clearly established at 5.7~nm (dashed vertical line), the actual separation in our bi-layer sample. This coupling is also evident in the calculated electron density and doping potential (Fig.~3d) where the single layer densities and potentials are seen to overlap. In an infinite stack of layers (Fig.~3e) the overlap between the electron density (and the potential) of each layer is further enhanced, supporting the finding of strong 3D inter-layer coupling in our multi-layer sample.
	
		In conclusion, our bottom-up approach to doping in Ge is capable of producing high electron densities (10$^{19}$ to 10$^{20}$~cm$^{-3}$) and low-resistivity (10$^{-4}~\Omega\cdot$cm) metallic conductors of precisely defined thickness. As such, this technology has immediate relevance in electronics, photonics, and plasmonics, towards the development of high mobility transistors, industrially viable Si-integrated lasers, and mid-IR plasmonics bio-sensors, respectively. Finally, tunable doping at high densities also provides an ideal test-bed to clarify open questions on doping-induced bottom-up superconductivity in group-IV semiconductors, as recently proposed by Shim and Tahan.\cite{25}
\\
\section*{Methods}
\textbf{Sample preparation.} All samples were fabricated in a customized ultra-high vacuum system (base pressure $< 5\times10^{-11}$~mbar) comprising a MBE system (MBE Komponenten) for Ge deposition and an additional chamber for surface preparation and PH$_{3}$ dosing. The chambers are connected via a UHV transfer tube. Ge(001) samples $2.5\times10$ mm$^2$ in size were cleaved from a Sb doped Ge(001) 4 inch wafer (resistivity of 1--10~$\Omega\cdot$cm). Atomically flat, clean, and defect-free surfaces Ge(001) surfaces were prepared for all samples with the method detailed in Ref \onlinecite{26}. In brief, an $ex$-$situ$ wet chemical treatment using HCl:H$_{2}$O (36:100) and H$_{2}$O$_{2}$:H$_{2}$O (7:100) is used to alternately strip and reform a germanium oxide passivation layer. This is subsequently removed $in$-$situ$ by a flash-anneal at 760~$\degree$C, followed by a 25-nm Ge buffer layer growth by MBE at a rate of 0.015~nm/s and sample temperature of 400~$\degree$C. Prior to the first doping cycle step, the surface is flattened by a final thermal anneal at 760~$\degree$C. For P doping, all samples were saturation-dosed at room temperature with PH$_{3}$ gas backfilling the UHV chamber at a pressure of  $\approx5\times10^{-11}$~mbar via a leak valve. Thermal incorporation of P atoms into the Ge surface was obtained by increasing the sample temperature from room temperature to 400~$\degree$C at a rate of 1~$\degree$C/s. Epitaxial growth of Ge spacers 5.7-nm-thick was performed by MBE with the following growth temperature sequence. The first $\approx0.7$~nm of the Ge spacer are deposited at 400 $\degree$C, the following $\approx2$~nm at 250~$\degree$C, and the final 3~nm, again, at 400~$\degree$C. Between the first and second step of the sequence, the growth is interrupted to allow for sample cool-down at a rate of 1~$\degree$C/s. This sequence was engineered to minimize dopant diffusion and segregation whilst retaining a low-roughness surface at each doping cycle.\cite{27} The deposition process ends for all samples with a 30-nm-thick Ge cap layer obtained by extending the duration of the third step in the growth temperature sequence.

\textbf{Pulsed laser atom probe tomography.} After removal from UHV, all samples were cleaved and one portion was used to produce atom probe compatible specimens. All samples were coated with protective amorphous films of 25--60~nm Cr/20~nm Pt using a broad ion beam sputter system (Gatan 682 Precision Etching and Coating system). The coated samples were then transferred to a focused ion beam (FIB)/scanning electron microscope (SEM) dual-beam system (FEI Nova 200 Nanolab, Hillsboro, OR). Atom probe compatible needle specimens were prepared in the dual beam system using site specific lift-out techniques, then mounted to Si microtip posts (CAMECA Atom Probe Technology Center, Madison, WI) and finally annular milled. Pulsed laser atom probe tomography was performed at the University of North Texas Center for Advanced Research and Technology (CART) using a Local Electrode Atom Probe (LEAP) 3000X HR (CAMECA Atom Probe Technology Center, Madison, WI) laser pulsed local electrode atom probe with a reflectron lens. Samples were analysed at a base temperature of 30--50~K in laser pulsed field evaporation mode using a pulsed laser with a wavelength of 532~nm, pulse width of 10~ps, applied at a pulse frequency of 16~ kHz, laser energy of 0.2--0.3~nJ and an evaporation rate of 0.001--0.005 ions/pulse. The resulting tomographic atom probe data was analysed using the atom probe reconstruction software, IVAS 3.6.1 (CAMECA Atom Probe Technology Center, Madison, WI). Reconstructions were correlated with Transmission Electron Microscopy and Secondary Ion Mass Spectrometry data (see Supplementary Section 1 for more details).

\textbf{Electrical characterization.} Trench-isolated Hall bars structures to investigate the electrical properties of the doped layers were defined by a CHF$_{3}$/CF$_{4}$ based dry etch with thermally evaporated Al Ohmics connecting in parallel all multiple P layers of the doped stack. Electrical characterization at 4.2~K was performed using a dipstick in liquid helium equipped with a superconducting magnet providing a perpendicular magnetic field up to 2~T. Characterization of the device at lower temperatures between 0.2 and 5~K was performed in a cryogen-free dilution refrigerator, equipped with a triple axis vector magnet system (Leiden Cryogenics B.V.). This enables independent control of both perpendicular and parallel components of the magnetic field, with respect to the average dopant plane, and allows for magnetic field rotations to be performed at fixed field. The vector magnet was critical to performing reliable WL measurements to extract the 2D layer thickness since the large anisotropy in $\Delta\sigma^{WL}$ requires alignment of $\vec{B}$ to better than 0.5 degrees with respect to the doping plane. For all measurements we used a four-probe setup using standard low frequency lock-in techniques and low injection currents ($\approx1$~nA) to measure simultaneously the magnetic field dependence of the longitudinal $\rho_{xx}$ and transverse $\rho_{xy}$  components of the resistivity tensor, where $x$ and $y$ are, respectively, the directions parallel or perpendicular to the current flow in the Hall Bar. The longitudinal $\sigma_{xx}$  and transverse $\sigma_{xy}$ (Hall) conductivity were calculated from the measured resistivities via tensor inversion. 

\textbf{Density functional theory calculations.} DFT calculations on stacked phosphorus dopant layers in germanium were conducted using the SIESTA software\cite{28} and methods described for single Ge:P layers in Ref. \onlinecite{24}. The DFT equations were solved using an atom-centered, double-numerical-plus-polarization (DNP) basis set and the local density approximation (LDA) with empirical on-site (+U) correction. The single and double dopant layer structures (Fig.~3a,~b and Fig.~3c,~d) were represented using highly elongated germanium super cells of 300 atomic layers which is sufficient to separate the dopant layers from their periodic repeats. For the repeated dopant layer stack (Fig.~3e) a much smaller unit cell of 40 layers was used in order to match the experimental layer separation of 5.7~nm. Phosphorus densities of $6.3\times10^{13}$~cm$^{-2}$ and $3.1\times10^{13}$~cm$^{-2}$ in the dopant plane were represented using the mixed-atom approach described in Ref. \onlinecite{24}. All dopant atoms in the calculations are confined to a single atomic plane.

\section*{Acknowledgements}
G.S. and G.C. acknowledge support from the Australian Research Council (project number DP130100403). M.Y.S acknowledges an Australian Research Council Laureate Fellowship and support from the Australian Research Council Centre of Excellence for Quantum Computation and Communication Technology (project no. CE110001027). A.R.H. acknowledges support from the ARC DP scheme, APF, and DORA awards.
\\
\section*{Contributions}
G.S. and W. M. K. fabricated the samples. D. L. J. characterised the samples by atom probe tomography. G. S. and L. A. Y. performed electrical measurements. G. S. and A. R. H. analysed the magnetotransport data. D. J. C., O. W, and N. A. M. carried out density functional theory calculations. G.S. planned the project. G.S. and O. W. prepared the manuscript with input from all authors.
\\
\section*{Additional information}
Supplementary information is available. Correspondence and request for materials should be addressed to G.S.
\\
\section*{Competing financial interests}
The authors declare no competing financial interests.\\

\onecolumngrid
\newpage

\section*{\large{Supplementary Information for\\''Bottom-up assembly of metallic germanium''}}
\section*{S1. Analysis of phosphorus depth profiles obtained by atom probe tomography}

Atom probe tomography was used to create three-dimensional tomographic reconstructions of the single layer, bi-layer, and multi-layer samples which are shown in Fig.~1c-e of the main text. In Figures S1a-c atom probe tomography data is integrated into phosphorus atomic concentration profiles and compared to concentration profiles obtained by Secondary Ion Mass Spectrometry (SIMS). This comparison highlights the superior depth resolution of atom probe tomography to characterise the sharp dopant profiles in our samples. The lower resolution in the SIMS data is due to ion beam mixing from the sputtering process, which is known to lead to artificial peak broadening. 
The atom probe phosphorus profile for the single layer sample in Fig.~S1a is slightly skewed to the left (i.e. towards the sample surface), which indicates that a degree of dopant segregation during growth is present. We fit this profile with an exponentially modified Gaussian (black line in Fig. S1a), which describes a diffusion-broadened profile with an exponential leading edge:\cite{1}
\begin{equation}
N(x) = A\sqrt{\frac{\pi}{2}}\frac{w}{\lambda}\textrm{exp}\left[\frac{\left(\mu-x\right)}{2}+\frac{1}{2}\left(\frac{w}{\lambda}\right)^{2}\right]\left[1-\textrm{erf} \frac{1}{\sqrt{2}} \left(\frac{\left(\mu-x\right)}{\lambda}+\frac{w}{\lambda}\right)\right]
\end{equation}

We use as fitting parameters the $1/e$ decay length $\lambda$, and the amplitude $A$, position $\mu$, and width $w$ of the deconvolution Gaussian. We find $\lambda= 0.71 \pm 0.03$~nm, which corresponds to a segregation length scale of approximately five atomic layer in the germanium crystal. This shows that the growth process induces very little dopant segregation and hence the doping profiles are sharp at the atomic level.

The peaks in the bi-layer and multilayer concentration profiles (Fig.~S1b,~c) appear symmetric. The small inter-layer spacing limits phosphorus segregation in these samples and the peak shape is primarily determined by dopant diffusion from the accumulated thermal budget of the repeated deposition process. Hence, in these samples the doping profiles are better fitted with Gaussian functions (black lines in Fig.~S1b and c). 

\begin{figure}
\center
\includegraphics[width=170mm]{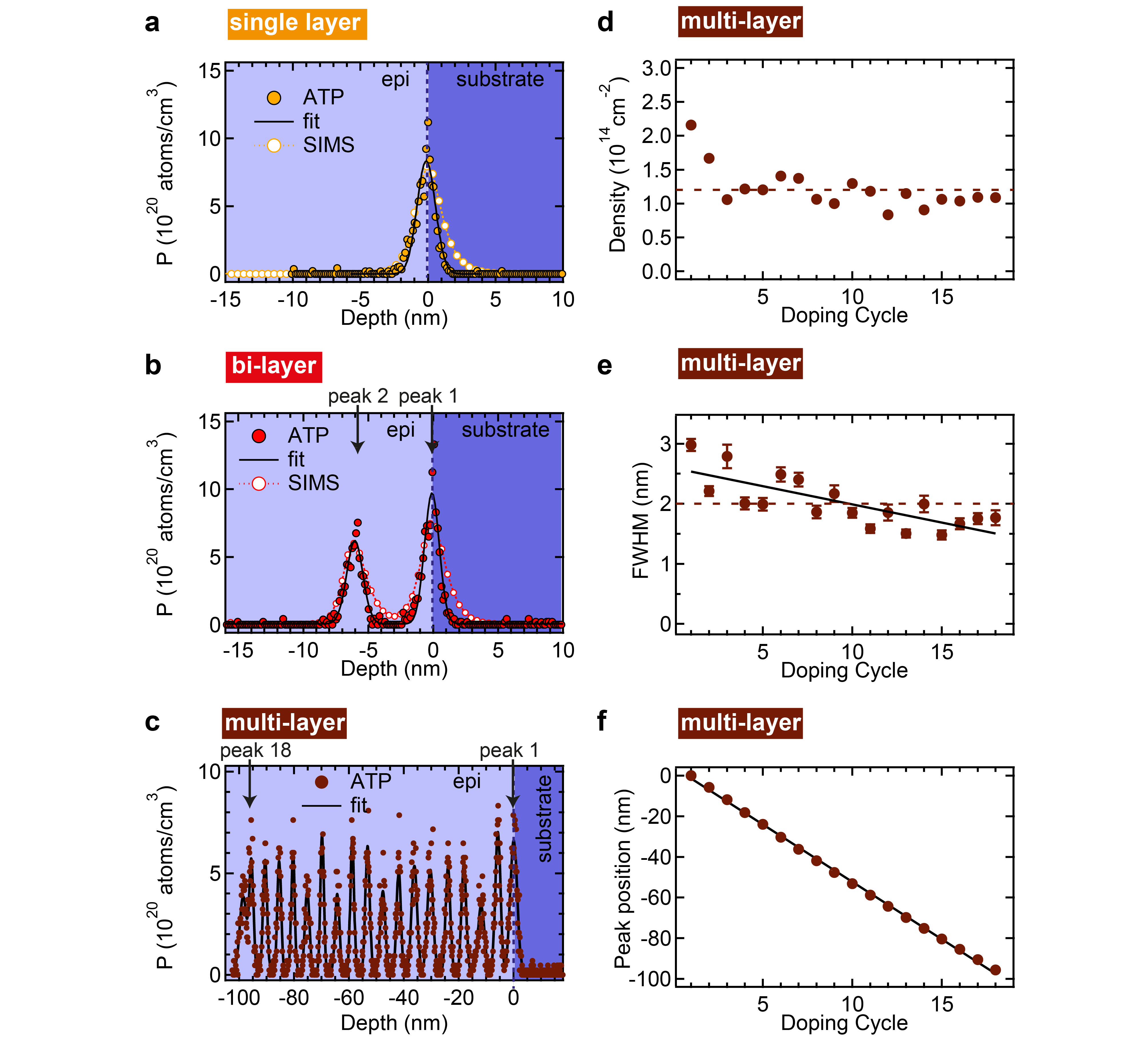}
\begin{flushleft}
\small  \label{fig1} \textbf{ Figure S1: Analysis of atom probe tomography results.} Phosphorus atomic concentration profiles from a (a) single layer, (b) bi-layer, and (c) multi-layer (18 layers) obtained by atom probe tomography (coloured circles) or SIMS (white circles) as a reference. Black lines are fit to the data. Analysis of individual peaks of the multi-layer sample showing: (d) planar doping density by integrating the area under each peak (dashed line is the average value); (e) full width at half maximum (FWHM) obtained by a Gaussian fit (dashed line is the average value);(f) peak position. Black lines in (e) and (f) are linear fits to the data from which we obtain the average FWHM broadening and inter-layer separation, respectively. In (d)-(f) the first doping cycle corresponds to the peak 1 in the concentration profile in (c).
\end{flushleft}
\end{figure}

\begin{table*}[htb]
\begin{minipage}{\linewidth}
 \centering 
    \renewcommand{\arraystretch}{1.25}
 \begin{tabular}{l c c c c c}
 \hline\hline
  & Single layer & Bi-layer(average) & Bi-layer (peak 2) & Bi-layer (peak 1) & Multi-layer (average)\\
   
 \hline
FWHM~(nm) & $1.41\pm0.05$ &$1.4\pm0.1$ & $1.47\pm0.05$ & $1.32\pm0.03$ & $2.0\pm0.4$\\
Area~($10^{14}$~cm$^{-2}$) & 1.44 &$1.2\pm0.3$ & 1.47 & 1 & $1.2\pm0.3$\\
 \hline 
\tabularnewline
\end{tabular}

\textbf{Table S1.} Analysis of phosphorus concentration profiles for single layer, bi-layer, and multilayer sample shown in Fig. S1a-c.\label{TableS1l}
\end{minipage}
\end{table*}

Table S1 summarises for all samples the key parameters reported in the main text. These parameters are the peak full width at half maximum (FWHM) and the area under the peak, obtained by numerical integration. In the single layer the FHWM is indicative of the dopant layer thickness $h$. We assume for FHWM in the single layer the same percentage error (3.6\%) of the deconvoluted Gaussian FWHM obtained from fitting Eq. (1) to the data. In the bi-layer, the thickness is estimated by adding the finite average peak width to the interlayer separation, $h = d+\mbox{FHWM} = 7.3 \pm 0.2$~nm. For all samples, the area under each peak is indicative of the planar P density per layer (n$_{P}$). The values achieved are comparable to previous studies of single monolayer doped Ge:P samples.\cite{2}

In the multilayer sample, the average peak FWHM is increased to approximately 2~nm. To investigate the origin of this broadening, the FWHM for each layer are shown in Fig. S1e. The data shows that the width progressively decreases from the first deposited layer to the last. This can be understood in terms of an accumulated thermal budged. The first deposited layer is the broadest because it undergoes a small amount of thermal diffusion during the annealing and growth step of each subsequent doping cycle. The last deposited layer is the sharpest because experiences only a single thermal processing step. A linear fit of FWHM versus Doping Cycle quantifies this broadening as approximately 0.06~nm per doping cycle. Despite this broadening, the layers are still distinct and their separation preserved. This is illustrated in Fig. S1e, where the linear increase in the peak position with the doping cycle yields an inter-layer separation value of  $5.65 \pm 0.05$~nm, as reported in the main text.

\section*{S2. Quantitative analysis of the weak localisation feature.}

In this section we provide further details into our measurements of the magnetoconductivity $\sigma_{xx}(\vec{B})$ with orientation of the vector magnetic field perpendicular ($B_{\perp}$) or parallel ($B_{\lVert}$) with respect to the 2D doping layers. We perform a quantitative analysis of the weak localisation correction in perpendicular magnetic field to obtain the relevant timescales associated with electron transport, indicating strong electronic coupling between stacked dopant layers. Throughout the range of temperatures investigated, electron transport is in the diffusive regime with very short measured elastic scattering times $\tau_{e}$ of $0.9\times10^{-14}$, $1.0\times10^{-14}$, and $1.8\times10^{-14}$~s in the single layer, bi-layer, and multi-layer sample, respectively. 

Figure S2a, b, and c compare the conductivity  $\sigma_{xx}$ as a function of $B_{\perp}$ and $B_{\lVert}$ (black and grey curves, respectively) for the single layer, bi-layer, and multi-layer, respectively. In the single layer (Fig.~S2a) weak localisation due to phase coherent backscattering reduces significantly the conductivity at zero magnetic field. The characteristic signature of weak localisation is a peaked positive magnetoconductance in perpendicular field, as $B_{\perp}$ threads flux through closed loops of particle backscattering trajectories and kills the weak localisation feature. In contrast, when the magnetic field is parallel to the 2D doping plane, very little flux is threaded through the particle trajectories, so almost no suppression of weak localisation is observed, and, consequently, the magnetic field dependence is weak. At $B$ = 1~Tesla (circles in Fig.~S2a) a strong anisotropy is established. This anisotropy is also evident in the polar plot in Fig 2b of the main text. In line with previous weak localisation measurements on Ge:P single layers,\cite{3} we observe in parallel magnetic field a small negative weak anti-localization correction, due the local moments in the presence of strong Coulomb interactions.
		
In the bi-layer sample (Fig. S2b), the perpendicular field dependence resembles that of the single layer sample. In the parallel direction, there is now significant field dependence because $B_{\lVert}$ threads flux through particle backscattering trajectories formed in-between layers due to strong coupling. In this sample, the anisotropy at 1~Tesla is still present but diminished. In the multi-layer sample (Fig.~S2c), the perpendicular and parallel magnetoconductance are seen to be almost overlapping, and hence the WL correction to the magnetoconductivity is nearly isotropic.

In Figure S2d we show the perpendicular field magnetoconductivity per layer $\sigma_{xx}(B_{\perp})/N$ with $N$ = 1,  2, and 18 in the single-layer, bi-layer and multilayer sample respectively. The measurements are over an extended range of magnetic fields to compare directly the magnitude of the WL correction across the three samples. We note a progressive suppression of the WL feature around zero magnetic field as the number of layers is increased, confirming the evolution of the electronic system from 2D to 3D. Note that if the layers were fully decoupled, all samples would show a similar dip in the magnetoconductance per layer. Associated with the dimensional cross-over is also the approximately twofold increase in electron mobility from the single-layer to the multi-layer. This is due to the population of 3D states extended in the vertical direction with reduced scattering from impurities within the doping planes.

\begin{figure*}
\center
\includegraphics[width=85mm]{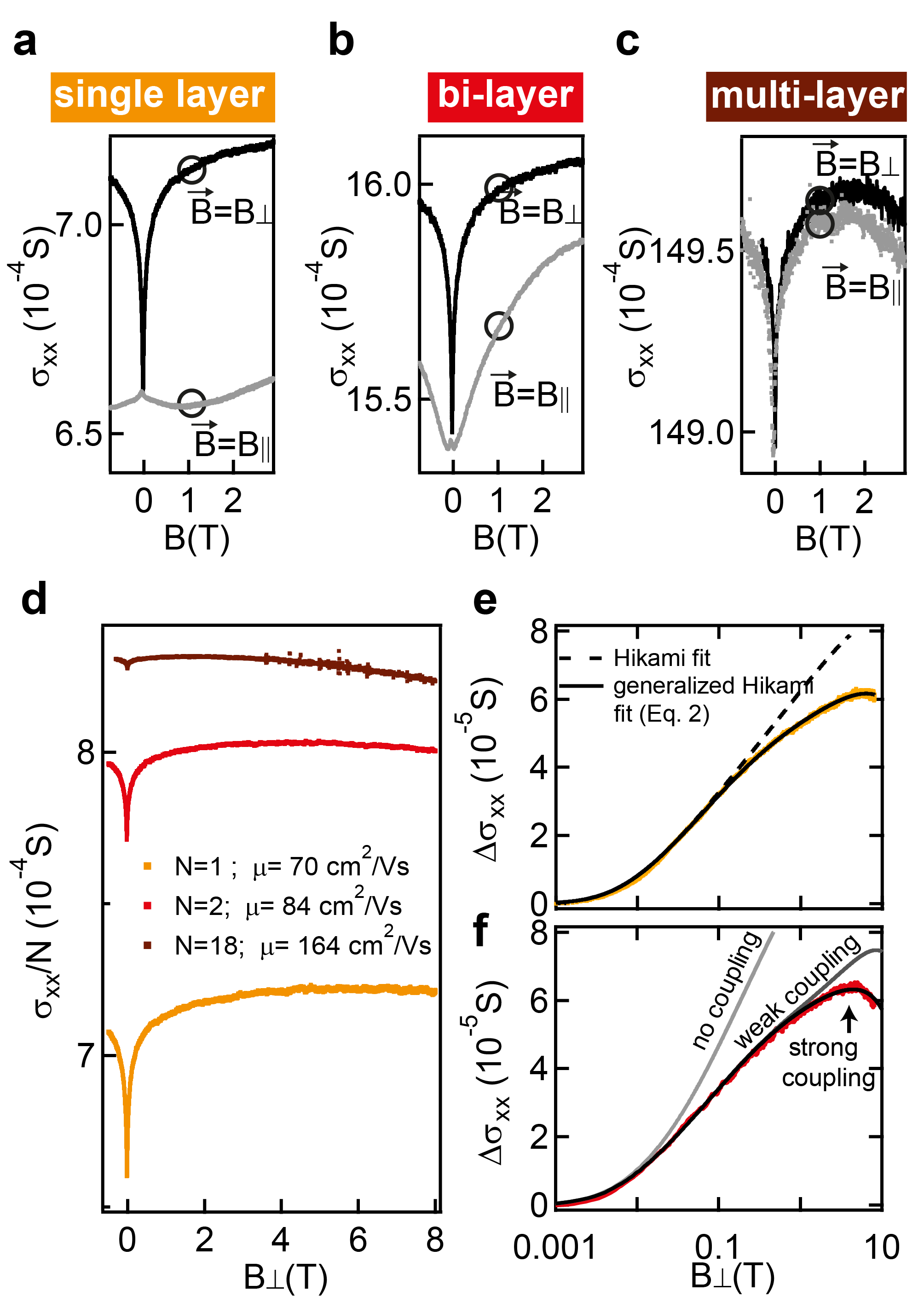}
\begin{flushleft}
\small\label{fig2}\textbf{Figure S2: Magnetoconductivity in perpendicular and parallel magnetic field.} Conductivity as a function of perpendicular or parallel magnetic field (black and grey lines, respectively) for a (a) single layer, (b) bi-layer, and (c) multi-layer (18 layers) at a temperature of 200~mK. (d) Conductivity per layer as a function of perpendicular magnetic field for the single layer, bi-layer, and multi-layer sample obtained by dividing the black curves in (a), (b), and (c) by 1, 2, and 18, respectively. (e) Weak localization correction to the magnetoconductivity for the single layer sample and theoretical fittings (black curves) obtained with the models described in the text. (f) Weak localization correction to the magnetoconductivity for the bi-layer sample and theoretical fits (grey and black curves) obtained for different electronic coupling regimes. Models are described in the text. 
\end{flushleft}
\end{figure*}

The relevant timescales of electron transport in the single layer and an insight into electronic coupling in the bi-layer sample are obtained by fitting of the WL feature in perpendicular field. For all fits we have subtracted the classical correction to the Drude conductivity $\sigma_{D}(\mu B)^2$. In the single layer (Fig.~S2e), the magnetoconductivity $\Delta\sigma_{xx}=\sigma_{xx}(B_{\perp})-\sigma_{xx}(0)$ is fitted to a phenomenological generalized Hikami-Larkin-Nagaoka expression for the quantum interference correction to the conductivity:\cite{3} 

\begin{equation}
\Delta\sigma_{xx}(B_{\perp})=\frac{\alpha e^2}{2 \pi^2\hbar}\left[F\left(\frac{\hbar}{4eD\tau_{\varphi}B_{\perp}}\right)-F\left(\frac{\hbar}{4eD\tau_{e}B_{\perp}}\right)\right]-\frac{\beta e^2}{2 \pi^2\hbar}F\left(\frac{\hbar}{4eD\tau_{s}B_{\perp}}\right)
\end{equation}				

In this expression $\alpha$ and $\beta$ are positive constants close to unity, $F(x)=\digamma(0.5+x)-ln⁡(x)$ with $\digamma(x)$ being the digamma function, $D$ is the two-dimensional diffusion constant, $\tau_{s}$ and $\tau_{\varphi}$ are the timescales characterising quasielastic spin scattering and dephasing, respectively. Excellent agreement between fit (black curve) and experimental data is obtained over the entire range of magnetic fields using values for the fitting parameters $\alpha$, $\beta$, $\tau_{s}$, and $\tau_{\varphi}$ listed in table S2. Note that fitting with a simple Hikami model (Fig.~S2e, dashed line), recovered by assuming $\beta$ = 0 in Eq. (2), neglects quasi elastic spin-scattering and fails to describe the data over the entire range of the magnetic field. 
In the bi-layer sample (Fig.~S2f), excellent agreement between fit (black curve) and experimental data is also obtained using Eq. (2), i.e. assuming that the two layers are coupled such that they effectively behave as a single coherent layer. To gain further insight into the electronic coupling in the bilayer, we also present in Fig. S2f theoretical fits (grey curves) to the expression: 
								
\begin{equation}
\Delta\sigma_{xx}(B_{\perp})=\frac{\alpha e^2}{2 \pi^2\hbar}\left[F\left(\frac{\hbar}{4eD\tau_{\varphi}B_{\perp}}\right)+F\left(\frac{\hbar}{4eDB_{\perp}}\frac{1+2\frac{\tau_{\varphi}}{\tau_{t}}}{\tau_{\varphi}}\right)\right]-\frac{\beta e^2}{2 \pi^2\hbar}F\left(\frac{\hbar}{4eD\tau_{s}B_{\perp}}\right)
\end{equation}				

This expression describes tunnel-coupled double quantum wells with a characteristic time $\tau_{s}$  as in Ref \onlinecite{4}, and is generalized to include quasi elastic spin scattering as in Eq. (2). For strong tunnel coupling ($\tau_{t}\leq\tau_{e}\ll\tau_{\varphi}$), Eq.~(3) reproduces the black curve in Fig~S2f, recovering the single layer limit. When the layers are weakly coupled (grey curve in Fig. S2f; $\tau_{t}/\tau_{e} = 10$) or fully de-coupled layers (light grey curve in Fig. S2f; $\tau_{t}/\tau_{e} = 1000$) the theoretical fits fail to fully reproduce the experimental data. Overall, this quantitative analysis confirms that electron transport in the bi-layer is in the strong coupling regime. As described in the main text, this regime is characterised by the coherent tunnelling of electrons between layers over a timescale comparable to the scattering off dopants within each layer.

\begin{table}[htb]
\begin{minipage}{\linewidth}
 \centering 
    \renewcommand{\arraystretch}{1.25}
 \begin{tabular}{l c c}
 \hline\hline
  & Single layer & Bi-layer\\
    \hline
$\alpha$ & 	$1.075\pm0.004$ & $1.056\pm0.007$\\
$\beta$ & 	$0.617\pm0.001$ & $0.693\pm0.001$\\
$\tau_e$ & 	$0.0097$ & $0.01$\\ 
$\tau_{\varphi}$ & $25.0\pm0.2$ & $30.0\pm0.04$\\
$\tau_s$ & $0.61\pm0.01$ & $0.34\pm0.01$\\
$\tau_t$ &  & $\leq0.01$\\       
\hline 
\tabularnewline
\end{tabular}

\end{minipage}
\begin{flushleft}
\textbf{Table S2.} Transport parameters for the single layer and bi-layer sample. $\tau_{e}$ was obtained directly by resistivity and Hall effect measurements. All other parameters were obtained by fitting the weak localisation feature in the magnetoconductance at a temperature of 200 mK. 
\end{flushleft}
\end{table}

\section*{S3. Fitting procedure to extract layer thickness from the angular dependence of the weak localisation corrections to the magnetoconductivity}

The thickness $h$ of the samples showing two-dimensional character, i.e. the single and bi-layer, is extracted by a quantitative analysis of the angular dependence of the weak localisation corrections to the magnetoconductivity $\Delta\sigma^{WL}(\vartheta)=\sigma(|\vec{B}|, \vartheta)$ measured at $|\vec{B}| = 1$~Tesla. As reported for 2DEGs hosted in other materials systems, such as Si MOSFETs,\cite{5} Si:P delta doped layers,\cite{6} and graphene,\cite{7} the effect of a parallel magnetic field is to enhance the electron dephasing rate by

\begin{equation}
\tau_{\varphi *}^{-1} \rightarrow \tau_{\varphi}^{-1}+\tau_{B_{\lVert}}^{-1}
\end{equation}

According to Ref. \onlinecite{5}, the additional dephasing rate in parallel magnetic field is given by

\begin{equation}
\tau_{B_{\lVert}}^{-1}=\left(\frac{e}{\hbar}\right)^2\frac{\sqrt{\pi}l_{e}}{\tau_{e}}Z^2RB_{\lVert}^2
\end{equation}

where $ l_{e}$ is the mean free path (obtained from resistivity and Hall effect measurements), $Z$ is the root mean square amplitude of the surface describing the two-dimensional electron gas, and $R$ is correlation length over surface height fluctuations. In our angular measurements performed at unitary magnetic field $B_{\lVert}^{2}=sin^{2}\vartheta$. As in Ref \onlinecite{6}, we assume that $Z$ and $R$ correspond to the thickness $h$ of the doping distribution probed electrically and the mean donor spacing within a layer, respectively. R is estimated from the atom probe tomography analysis as $1/\sqrt{n_{P}}$. We extract h by fitting $\Delta\sigma^{WL}(\vartheta)$ (black lines in Fig.~2a-c) to the phenomenological generalized Hikami-Larkin-Nagaoka expression given in Eq.~(2) with $B_{\perp}=cos\vartheta$ for $|\vec{B}| = 1$~Tesla and using the effective dephasing time $\tau_{\varphi *}$ defined in Eq. (4) instead of $\tau_{\varphi}$. $Z$ is used as a single fitting parameter and is linked to the effective dephasing time  $\tau_{\varphi*}$. As reported in the main text, we obtain thicknesses of $1.49 \pm 0.03$~nm and $6.17 \pm 0.05$~nm in the single and bi-layer sample, respectively. These values are both in agreement with the thickness obtained by atom probe tomography, confirming that the conducting regions in our samples match the extent of the stacked doping profiles.

\


\begin{thebibliography}{10}
\section*{References}

\bibitem{1}
Pillarisetty, R. Academic and industry research progress in germanium nanodevices. \textit{Nature} \textbf{479}, 324--328 (2011).

\bibitem{2}
Liang, D. \& Bowers, J.~E. Recent progress in lasers on silicon.\textit{Nature Photon.} \textbf{4}, 511--517 (2010).

\bibitem{3}
Liu, J.~F., Sun, X.~C., Camacho-Aguilera, R., Kimerling, L.~C. \& Michel, J. Ge-on-Si laser operating at room temperature.\textit{Opt. Lett.} \textbf{35}, 679--681 (2010).

\bibitem{4}
Soref, R. Mid-infrared photonics in silicon and germanium.\textit{Nature Photon.} \textbf{4}, 495--497 (2010).

\bibitem{5}
Soref, R., Hendrickson, J. \& Cleary, J.~W. Mid- to long-wavelength infrared plasmonic-photonics using heavily doped n-Ge/Ge and n-GeSn/GeSn heterostructures.\textit{Opt. Express} \textbf{20}, 3814--3824 (2012).

\bibitem{6}
Brotzmann, S. \& Bracht, H. Intrinsic and extrinsic diffusion of phosphorus, arsenic, and antimony in germanium.\textit{J. Appl. Phys.} \textbf{103}, 033508 (2008).

\bibitem{7}
Kamata, Y. High-k/Ge MOSFETs for future nanoelectronics.\textit{Mater. Today} \textbf{11}, 30--38 (2008).

\bibitem{8}
Dutt, B. \textit{et al.} Roadmap to an Efficient Germanium-on-Silicon Laser: Strain vs. n-Type Doping.\textit{Ieee Photonics J.} \textbf{4}, 2002--2009 (2012).

\bibitem{9}
Simoen, E. \& Vanhellemont, J. On the diffusion and activation of ion-implanted n-type dopants in germanium.\textit{J. Appl. Phys.} \textbf{106}, 103516 (2009).

\bibitem{10}
Chroneos, A. \& Bracht, H. Diffusion of n-type dopants in germanium.\emph{ Appl. Phys. Rev.} \textbf{1}, 011301 (2014).

\bibitem{11}
Murota, J., Sakuraba, M. \& Tillack, B. Atomically controlled processing for group IV semiconductors by chemical vapor deposition.\textit{Jap. J. appl. Phys.} \textbf{45}, 6767--6785 (2006).

\bibitem{12}
Scappucci, G., Capellini, G., Lee, W.~C.~T. \& Simmons, M.~Y. Ultradense phosphorus in germanium delta-doped layers.\textit{Appl. Phys. Lett.} \textbf{94}, 162106 (2009).

\bibitem{13}
Ho, J.~C. \textit{et al.} Controlled nanoscale doping of semiconductors via molecular monolayers.\textit{Nature Mater.} \textbf{7}, 62--67 (2008).

\bibitem{14}
Camacho-Aguilera, R.~E. \textit{et al.} An electrically pumped germanium laser.\textit{Opt. Express} \textbf{20}, 11316--11320 (2012).

\bibitem{15}
Scappucci, G. \textit{et al.} n-Type Doping of Germanium from Phosphine: Early Stages Resolved at the Atomic Level.\textit{Phys. Rev. Lett.} \textbf{109}, 076101 (2012).

\bibitem{16}
Mattoni, G., Klesse, W.~M., Capellini, G., Simmons, M.~Y. \& Scappucci, G. Phosphorus Molecules on Ge(001): A Playground for Controlled n-Doping of Germanium at High Densities.\textit{ACS Nano} \textbf{7}, 11310--11316 (2013).

\bibitem{17}
Spitzer, W.~G., Trumbore, F.~A. \& Logan, R.~A. Properties of Heavily Doped N-Type Germanium.\textit{J. Appl. Phys.} \textbf{32}, 1822 (1961).

\bibitem{18}
Weber, B. \textit{et al.} Ohm's Law Survives to the Atomic Scale.\textit{Science} \textbf{335}, 64--67 (2012).	

\bibitem{19}
Altshuler, B.~L., Aronov, A.~G. \& Lee, P.~A. Interaction Effects in Disordered Fermi Systems in 2 Dimensions.\textit{Phys. Rev. Lett.} \textbf{44}, 1288--1291 (1980).

\bibitem{20}
Hikami, S., Larkin, A.~I. \& Nagaoka, Y. Spin-Orbit Interaction and Magnetoresistance in the 2 Dimensional Random System.\textit{Prog. Theor. Phys.} \textbf{63}, 707--710 (1980).

\bibitem{21}
Shamim, S. \textit{et al.} Spontaneous Breaking of Time-Reversal Symmetry in Strongly Interacting Two-Dimensional Electron Layers in Silicon and Germanium.\textit{Phys. Rev. Lett.} \textbf{112}, 236602 (2014).

\bibitem{22}
Mathur, H. \& Baranger, H.~U. Random Berry phase magnetoresistance as a probe of interface roughness in Si MOSFET's.\textit{Phys. Rev. B} \textbf{64}, 235325 (2001).

\bibitem{23}
Sullivan, D.~F., Kane, B.~E. \& Thompson, P.~E. Weak localization thickness measurements of Si : P delta-layers.\textit{Appl. Phys. Lett.} \textbf{85}, 6362--6364 (2004).

\bibitem{24}
Carter, D.~J. \textit{et al.} Electronic structure of phosphorus and arsenic delta-doped germanium.\textit{Phys. Rev. B} \textbf{88}, 115203 (2013).


\bibitem{25}
Shim, Y.~P. \& Tahan, C. Bottom-up superconducting and Josephson junction devices inside a group-IV semiconductor.\textit{ Nature Commun.} \textbf{5}, 4225 (2014).

\bibitem{26}
Klesse, W.~M., Scappucci, G., Capellini, G. \& Simmons, M.~Y. Preparation of the Ge(001) surface towards fabrication of atomic-scale germanium devices.\textit{Nanotechnology} \textbf{22}, 145604 (2011).

\bibitem{27}
Scappucci, G., Capellini, G., Klesse, W.~M. \& Simmons, M.~Y. Phosphorus atomic layer doping of germanium by the stacking of multiple delta layers.\textit{Nanotechnology} \textbf{22}, 375203 (2011).

\bibitem{28}
Soler, J.~M. \textit{et al.} The Siesta method for ab initio order-N materials simulation.\textit{J. Phys.: Condens. Matter} \textbf{14}, 2745--2779 (2002).

\end{thebibliography}

\begin{thebibliography}{10}
\section*{References}

\bibitem{1}
Polley, C.~M. \textit{et al.}  Exploring the Limits of N-Type Ultra-Shallow Junction Formation. \textit{ACS Nano} \textbf{7},  5499--5505 (2013).

\bibitem{2}
Scappucci, G., Capellini, G. \& Simmons, M.~Y. Influence of encapsulation temperature on Ge:P delta-doped layers.\textit{Phys. Rev. B} \textbf{80}, 233202 (2009).

\bibitem{3}
Shamim, S. \textit{et al.} Spontaneous Breaking of Time-Reversal Symmetry in Strongly Interacting Two-Dimensional Electron Layers in Silicon and Germanium. \textit{Phys. Rev. Lett.} \textbf{112}, 236602 (2014).

\bibitem{4}
Raichev, O.~E. \& Vasilopoulos, P. Weak-localization corrections to the conductivity of double quantum wells. \textit{J. Phys.: Condens. Mat. } \textbf{12}, 589--600 (2000).

\bibitem{5}
Mathur, H. \& Baranger, H.~U. Random Berry phase magnetoresistance as a probe of interface roughness in Si MOSFET's.\textit{Phys. Rev. B} \textbf{64}, 235325 (2001).

\bibitem{6}
Sullivan, D.~F., Kane, B.~E. \& Thompson, P.~E. Weak localization thickness measurements of Si : P delta-layers. \textit{Appl. Phys. Lett.} \textbf{20}, 3814--3824 (2012).

\bibitem{7}
Lundeberg, M.~B. \& Folk, J.~A. Rippled Graphene in an In-Plane Magnetic Field: Effects of a Random Vector Potential. \textit{Phys. Rev. Lett.} \textbf{105}, 146804 (2010).


\end{thebibliography}
\end{document}